\documentstyle[aps,preprint,epsfig,floats]{revtex}

\def\slash#1{#1\!\!\!/}
\def\eqref#1{Eq.\ (\ref{#1})}
\def\figref#1{Fig.\ \ref{#1}}
\newcommand{\eqsref}[2]{Eqs.\ (\ref{#1}) -- (\ref{#2})}

\begin{document}
\setcounter{page}{0}
\def\footnoterule{\kern-3pt \hrule width\hsize \kern3pt}
\tighten

\title{On the Applicability of Weak-Coupling Results in High Density QCD}

\author{Krishna~Rajagopal\footnote{Email address: {\tt krishna@ctp.mit.edu}}
and Eugene~Shuster\footnote{Email address: {\tt eugeneus@mit.edu}}}

\address{Center for Theoretical Physics \\
%Laboratory for Nuclear Science \\
%and Department of Physics \\
Massachusetts Institute of Technology \\
Cambridge, MA 02139 \\
{~}}

\date{MIT-CTP-2969,~  hep-ph/0004074,~ April 7, 2000}
\maketitle

\thispagestyle{empty}

\begin{abstract}
Quark matter at asymptotically high baryon chemical potential is in a
color superconducting state characterized by a gap $\Delta$.
We demonstrate
that although present 
weak-coupling calculations of $\Delta$ are formally correct for
$\mu\rightarrow\infty$, the contributions which have to
this point been neglected are large enough
that present results can only be trusted for $\mu\gg \mu_c\sim 10^{8}$ MeV.
We make this argument by using
the gauge dependence of the present calculation as a
diagnostic tool.  It is known that the present calculation yields
a gauge invariant result for $\mu\rightarrow\infty$; we 
show, however, that the gauge dependence of this result 
only {\it begins} to decrease for $\mu\gtrsim\mu_c$, and conclude
that 
the result can certainly not be trusted for $\mu<\mu_c$.
In an appendix, we set up the 
calculation of the influence of the Meissner effect 
on the magnitude of the gap. This contribution to $\Delta$
is, however, much smaller than the neglected contributions whose
absence we detect via the resulting gauge dependence.
\end{abstract}

\vfill\eject

\section{Introduction}

The starting point for a description of matter at high baryon density
and low temperature is a Fermi sea of quarks. The important degrees of
freedom --- those whose fluctuations cost little free energy --- are
those involving quarks near the Fermi surface.  We know from the work of
Bardeen, Cooper, and Schrieffer \cite{BCS} that any attractive
interaction between the quarks, regardless how weak, makes the Fermi sea
unstable to the formation of a condensate of Cooper pairs. In QCD, the
interaction of two quarks whose colors are antisymmetric (the color
$\bar {\bf 3}_A$ channel) is attractive. (The attractiveness of this
interaction can be seen from single-gluon exchange, as is relevant at
short distances, or via counting strings or analyzing the instanton
induced coupling, as may be relevant at longer distances.)  We therefore
expect that under any circumstance in which cold dense quark matter is
present, it will be in a color superconducting
phase \cite{Barrois1,Barrois2,BailinLove,ARW1,RSSV}.
The one caveat is that this conclusion is known to be false if
the number of colors is $N_c=\infty$\cite{DGR}.
Recent work \cite{ShusterSon,ParkRhoWirzbaZahed} indicates that
quark matter is in a color superconducting phase
for $N_c$ less than of order thousands, and in this paper we only discuss
QCD with $N_c=3$.

We now know much about the symmetries and physical properties of color
superconducting quark matter.  The dominant condensate in QCD with two
flavors of quarks is in the color $\bar {\bf 3}_A$ channel, breaking
$SU(3)_{\rm color}\rightarrow SU(2)$, and is a flavor
singlet\cite{Barrois1,Barrois2,BailinLove,ARW1,RSSV}. Quarks with two
of three colors have a gap in this 2SC phase, and five of eight
gluons get a mass via the Meissner effect. 
In QCD with three flavors of quarks, the Cooper
pairs cannot be flavor singlets, and flavor symmetries are necessarily
broken. The symmetries of the phase which results have been analyzed in
Ref. \cite{CFL}, and were in fact first analyzed in a different
(zero density) context in Ref. \cite{SrednickiSusskind}. 
The dominant condensate locks color and flavor
symmetries, leaving an unbroken global symmetry under simultaneous
$SU(3)$ transformations of color, left-flavor, and right-flavor. In this
CFL phase, all nine quarks have a gap and all eight gluons have a
mass\cite{CFL}. 
Chiral symmetry is spontaneously broken, as is baryon number, and
there are consequently nine massless Goldstone bosons \cite{CFL}.  
Matter in the CFL phase is therefore similar in many respects to 
superfluid hypernuclear matter \cite{CFL,SWcont,Unlocking,SWUnlocking}.
The fact that color superconducting phases always feature either
chiral symmetry breaking (as in the CFL phase) or some quarks which
remain gapless (as in the 2SC phase) may be understood as a consequence
of imposing 't Hooft's anomaly matching criterion \cite{Sannino}.
The
first order phase transition between the CFL and 2SC phases has been
analyzed in detail\cite{Unlocking,SWUnlocking,Gapless}, but all that will
concern us below is that any finite strange quark mass is unimportant at
large enough $\mu$, and quark matter is therefore in the CFL phase
at asymptotically large $\mu$.  

Much recent work has resulted in two classes of
estimates of the magnitude of $\Delta$, the gap in the density of
quasiparticle states in the superconducting phase.  The first class of
estimates are done within the context of models whose parameters are
chosen to give reasonable vacuum physics. Examples include analyses in
which the interaction between quarks is replaced simply by four-fermion
interactions with the quantum numbers of the instanton
interaction\cite{ARW1,RSSV,BergesRajagopal} or of single-gluon
exchange\cite{CFL,Unlocking} and more sophisticated analyses done using
instanton liquid
models\cite{CarterDiakonov,RSSV2}. Renormalization
group analyses have also been used to explore the space of all possible
four-fermion interactions allowed by the symmetries of
QCD\cite{Evans1,SWRG}. These methods yield results which are in
qualitative agreement: the gaps range from several tens of MeV up to as
much as about 100 MeV and the corresponding critical temperatures, above
which the superconducting condensates vanish, can be as large as about
50 MeV. 

The second class of estimates uses $\mu\rightarrow\infty$ physics 
as a guide. At
asymptotically large $\mu$, models with short range interactions are
bound to fail, because the dominant interaction is due to the long-range
magnetic interaction coming from single-gluon
exchange\cite{PisarskiRischke1,Son}. The collinear 
infrared divergence
in small-angle scattering via single-gluon exchange results in a gap
which is parametrically larger at $\mu\rightarrow\infty$ than it would
be for any point-like four-fermion interaction\cite{Barrois2}. Son
showed \cite{Son} that this collinear divergence is regulated by
Landau damping (dynamical screening) 
and that as a consequence, the parametric dependence
of the gap in the limit in which the QCD coupling $g\rightarrow 0$ is
\begin{equation}
\frac{\Delta}{\mu} \sim \frac{1}{g^5} 
\exp\left(-\,\frac{3\pi^2}{\sqrt{2} g}\right)\ ,
\label{sonequation}
\end{equation}
which
is more easily seen as an expansion in $g$ when rewritten as
\begin{equation}
\ln\left(\frac{\Delta}{\mu}\right) = -\frac{3\pi^2}{\sqrt{2}}\frac{1}{g}
-5\ln g + f(g) \ .
\label{expansion}
\end{equation}
This equation should be viewed as a definition of $f(g)$, which will
include a term which is constant for $g\rightarrow 0$ and 
terms which vanish for $g\rightarrow 0$.  The result (\ref{sonequation})
has been 
confirmed using a variety of 
methods\cite{SW,PisarskiRischke3,Hong,HMSW,rockefeller,HsuSchwetz}, 
and several
estimates of $\lim_{g\rightarrow 0}f(g)$ exist in the literature.
For example, Schaefer and Wilczek find\cite{SW,Schaefer}
\begin{equation}
\lim_{g\rightarrow 0}f(g) \sim \ln\left[ 2^{-1/3} 256 \pi^4
\left(\frac{2}{3}\right)^{5/2} \right] = 8.88
\label{SWc}
\end{equation}
in the CFL phase (see also 
Ref. \cite{PisarskiRischke3}),  
and Brown, Liu, and Ren\cite{rockefeller} 
find a result for $\lim_{g\rightarrow 0}f(g)$ which 
is smaller by $(\pi^2 +4)/8 - \ln 2 = 1.04$.
If this asymptotic expression is applied by taking $g=g(\mu)$ from the
perturbative QCD $\beta$-function (with $\Lambda_{\rm QCD}=200$ MeV),
evaluating $\Delta$ at
$\mu\sim 500$ MeV yields gaps in
rough agreement with the estimates based on zero-density phenomenology.  

The central purpose of this paper is to demonstrate that
this nice agreement must at present be seen as coincidental,
because present estimates for $f$ are demonstrably uncontrolled
for $g>g_c\sim 0.8$, corresponding to $\mu<\mu_c$ with $\mu_c\sim 10^{8}$
or higher. 

The weak-coupling calculations are derived from analyses
(done using varying approximations) of the one-loop
Schwinger-Dyson equation without vertex correction, and 
(with one exception)
yield gauge dependent results.
However,
Schaefer and Wilczek argue that the result for 
$\lim_{g\rightarrow 0}f(g)$ in such a calculation 
is gauge invariant.  The one calculation which
is gauge invariant throughout
is the calculation of $T_c$ (and 
hence $\Delta$ since the BCS relation $T_c=0.57 \Delta$ 
holds \cite{PisarskiRischke3})
done by Brown, Liu, and Ren\cite{rockefeller}.
As in other calculations, however, these authors neglect 
vertex corrections.  Our purpose is to
{\it use} the fact that our calculation (like most) is 
gauge dependent, and only gauge invariant for
$g\rightarrow 0$, to estimate the $g$ above which vertex
corrections,  left out of {\it all} calculations, cannot
be neglected.

We begin by sketching the 
derivation of the one-loop Schwinger-Dyson equation for $\Delta$,
making as few approximations as we can.  We solve
the resulting gap equation numerically in several different
gauges.  Our results are (yet one more) confirmation
of (\ref{sonequation}). Furthermore, 
we do find evidence that the gauge dependence of $f$ decreases
for $g\rightarrow 0$.  However, this decrease only begins
to set in for $g\lesssim 0.8$.  This implies that the
contributions to $\Delta$ which have been neglected --- like
those arising from vertex corrections --- only become subleading
for $g\ll g_c\sim 0.8$.  If we translate $g_c$ to $\mu_c$ 
by assuming $g$ should be taken as $g(\mu)$, this
corresponds to $\mu_c\sim 10^{8}$ MeV.  Recent
work \cite{BBSunpub} shows that $g$ should be evaluated
at a much lower ($g$-dependent) scale than $\mu$.  This means
that the condition $g<g_c\sim 0.8$ would translate into $\mu>\mu_c$
with $\mu_c$ orders of magnitude larger than $\mu_c\sim 10^{8}$ MeV.

The original purpose of our investigation was 
to do a self-consistent calculation of the influence of
the Meissner effect on the magnitude of the gap in the 
CFL phase. In the
presence of a condensate, the gluon propagator is modified: some gluons
get a mass.  In the CFL phase, {\it all} gluons get a mass, and this
makes a calculation based on perturbative single-gluon exchange a
self-consistent and complete description of the physics at
asymptotically large $\mu$, with no remaining infrared problems.  (In
the 2SC phase, in contrast, the calculation of $\Delta$ leaves
unanswered any questions about the non-Abelian
infrared physics of the three gluons
left unscreened by the condensate.)  We felt that this motivation warranted
a self-consistent calculation in which we calculate the gap
using a Schwinger-Dyson equation in which the gluon
propagator is modified not only by the presence of the Fermi sea (Debye
mass, Landau damping) but is also affected by the condensate (the
Meissner effect).  We set this calculation up in an
appendix. Previous work, beginning with that 
of Ref. \cite{Son}, shows that the form 
of \eqref{sonequation}
is unmodified by including the Meissner effect, but $f(g)$ is affected. Our
preliminary results suggest that the changes in $f(g)$ are small, as
anticipated in Refs. \cite{Son,SW,HMSW,HsuSchwetz,ShovkovyWije,EHHS}.  
Indeed, the effects of physics left out of the
present analysis, which we have diagnosed via the gauge dependence of
$f(g)$, are much larger than those 
introduced by the Meissner effect at any $g$ we have
investigated.

\section{Deriving the gap equation}

In this section, we derive the gap equation 
for QCD with three 
massless flavors which is 
valid at asymptotically high densities. We follow Ref. \cite{SW},
but make fewer approximations.
Because our point is to stress the importance of
effects which we do {\it not} calculate, we will make
our assumptions and approximations very clear as we proceed.
In other words, since the lesson we learn from our results
is that they cannot yet be trusted, it is important to detail
carefully all points at which we leave something out.

We use the standard Nambu-Gorkov
formalism by defining an eight-component field
$\Psi=(\psi,\bar\psi^T)$. In this basis, the inverse quark propagator
takes the form
\begin{equation} \label{Sinv}
S^{-1}(k) = \left(\begin{array}{cc} \slash{k}+\mu\gamma_0 &
\bar\Delta \\ \Delta  & (\slash{k}-\mu\gamma_0)^T
\end{array}\right)
\end{equation}
where $\bar\Delta = \gamma_0\Delta^\dagger\gamma_0$. The color, flavor,
and Dirac indices are suppressed in the above expression.  The diagonal
blocks correspond to ordinary propagation and the off-diagonal blocks
reflect the possibility for ``anomalous propagation'' in the presence
of a diquark condensate.

We make the
following ansatz for the form of the gap 
matrix\cite{BailinLove,CFL,SW,PisarskiRischke2}:
\begin{eqnarray} \label{delta}
\Delta^{ab}_{ij}(k) & = (\lambda^A_I)^{ab}(\lambda^A_I)_{ij}
C\gamma_5 \left( {\Delta_1^A}(k_0)P_+(k) +
{\Delta_2^A}(k_0)P_-(k) \right) \nonumber \\ & +
(\lambda^S_J)^{ab}(\lambda^S_J)_{ij} C\gamma_5 \left( 
{\Delta_1^S}(k_0)P_+(k)
+{\Delta_2^S}(k_0)P_-(k) \right)
\end{eqnarray}
Here, $a, b = 1, 2, 3$ are color indices, $i, j
= 1, 2, 3$ are flavor indices, $\lambda^A_I$ are antisymmetric $U(3)$ 
color or flavor matrices with $I = 1, 2, 3$, 
and $\lambda^S_J$ are symmetric $U(3)$ color or flavor 
with $J = 1, \ldots, 6$, and the projection operators $P_\pm$
are defined as
\begin{eqnarray} \label{projectors}
P_+(k) = {{1+\vec\alpha\cdot\hat k}\over 2} \nonumber \\
P_-(k) = {{1-\vec\alpha\cdot\hat k}\over 2}
\end{eqnarray}
with $\vec\alpha = \gamma_0\vec\gamma$.

By making this ansatz, we are making several assumptions:
\begin{itemize}
\item
First, we have taken $\Delta_1^A$, $\Delta_2^A$,
$\Delta_1^S$, and  $\Delta_2^S$ to be functions of $k_0$ only.
All are in principle functions of both $k_0$ and $|\vec k|$, 
but we assume that they are dominated by $|\vec k|\sim \mu$.
This is a standard assumption, and although we do not expect
that relaxing this assumption would resolve the problems which
we diagnose below, this does belong on the list of potential cures.
\item
Second, we have explicitly separated the gaps which are antisymmetric
$\bar{\bf 3}_A$ in color and flavor from those which are symmetric 
${\bf 6}_S$ in color and flavor and, in both cases, we have assumed
that the favored channel is the one in which color and flavor 
rotations are locked.  The color and flavor structure of our
ansatz is thus precisely that first explored in Ref. \cite{CFL}, which
allows quarks of all three colors and all three flavors to pair.
Subsequent work \cite{Schaefer,ShovkovyWije,EHHS,Hong2} confirms 
that this is the favored condensate,
and we will not attempt to further generalize it here.  
\item
Third, we have assumed that the Cooper pairs in the condensate
have zero spin and orbital angular momentum.  This seems a safe
assumption in the CFL phase, where the dominant condensate, made
of Cooper pairs with
zero spin and orbital angular momentum, leaves no quarks ungapped.
\item
Fourth, we neglect $\bar\psi\psi$ condensates.  Since chiral
symmetry is broken in the CFL phase, these must be nonzero \cite{CFL}.
This applies to both color singlet
and color octet $\bar\psi\psi$ condensates\cite{Wetterich}.
Such condensates are small\cite{RSSV2,Schaefer},
however, and we expect that neglecting
them results in only a very small error in the magnitude
of the dominant diquark condensate.
\end{itemize}

The most important assumption we make is that we
obtain the gap by solving the
one-loop Schwinger-Dyson equation of the form
\begin{equation} \label{SDeq}
 S^{-1}(k)-S_0^{-1}(k) = ig^2 \int \frac{d^4q}{(2\pi)^4}
 \Gamma_\mu^a S(q)\Gamma_\nu^b D_{ab}^{\mu\nu}(k-q) \ ,
\end{equation}
using a medium-modified gluon propagator described below and
unmodified vertices 
\begin{equation} \label{vertex}
\Gamma_\mu^a = \left(\begin{array}{cc} \gamma_\mu\lambda^a/2 & 0 \\ 0 &
-(\gamma_\mu\lambda^a/2)^T \end{array}\right) .
\end{equation}
Here, $S_0$ is the bare fermion propagator with $\Delta=0$.
Note that we use a Minkowski metric unless stated otherwise.
We will demonstrate that our results
are completely uncontrolled for $g>g_c\sim 0.8$. This breakdown could
in principle reflect a failure of any of our assumptions.
We expect, however, that it
arises because contributions which have been truncated in
writing (\ref{SDeq}) are large for $g>g_c$.  That is, we expect
that this truncation (and not any of the simplifications 
introduced by
our ansatz for $\Delta$) is the most significant assumption 
we are making.

%\begin{figure} \centering
%\epsfig{file=graph2.ps,width=2in}%,bbllx=0,bblly=0,bburx=612,bbury=644}
%\caption{Schwinger-Dyson equation} \label{SDfig} 
%\end{figure}

We obtain four coupled gap
equations 
\begin{eqnarray} \label{4gapeq}
\Delta^A_{1,2}(k_0) = 
-{i\over 6} g^2 \int \frac{d^4q}{(2\pi)^4} {\rm Tr} \left[
P_\pm(k)\gamma_\mu \left( P_+(q) a_+(q) + P_-(q) a_-(q)\right) 
\gamma_\nu \right]
D^{\mu\nu}(k-q) \nonumber \\
\Delta^S_{1,2}(k_0) = 
-{i\over 6} g^2 \int \frac{d^4q}{(2\pi)^4} {\rm Tr} \left[
P_\pm(k)\gamma_\mu \left( P_+(q) b_+(q) + P_-(q) b_-(q)\right) 
\gamma_\nu \right]
D^{\mu\nu}(k-q)
\end{eqnarray}
where $P_\pm$ means $P_+$ in the $\Delta_1$ equation 
and $P_-$ in the $\Delta_2$ equation and where
\begin{eqnarray} \label{ab}
a_+(q) &=& {-{\Delta_2^S}(q_0)-{\Delta_2^A}(q_0) \over
q_0^2-(|\vec{q}|+\mu)^2- 4\left[{\Delta_2^A}(q_0)+2{\Delta_2^S}(q_0)\right]^2} 
\nonumber\\
&~~& + {\left[ {\Delta_2^A}(q_0)-{\Delta_2^S}(q_0) \right] \left[
-q_0^2+(|\vec{q}|+\mu)^2+(5{\Delta_2^A}(q_0)+7{\Delta_2^S}(q_0))
({\Delta_2^A}(q_0)+3{\Delta_2^S}(q_0))\right] \over 
\left[ q_0^2-(|\vec{q}|+\mu)^2-
({\Delta_2^A}(q_0)-{\Delta_2^S}(q_0))^2 \right] \left[ q_0^2-(|\vec{q}|+\mu)^2-
4({\Delta_2^A}(q_0)+2{\Delta_2^S}(q_0))^2 \right]}\nonumber\\
a_-(q) &=& {-{\Delta_1^S}(q_0)-{\Delta_1^A}(q_0) \over q_0^2-(|\vec{q}|-\mu)^2-
4\left[{\Delta_1^A}(q_0)+2{\Delta_1^S}(q_0)\right]^2}\nonumber\\
&~~& + {\left[ {\Delta_1^A}(q_0)-{\Delta_1^S}(q_0)
\right] \left[ -q_0^2+(|\vec{q}|-\mu)^2+(5{\Delta_1^A}(q_0)+7{\Delta_1^S}(q_0))
({\Delta_1^A}(q_0)+3{\Delta_1^S}(q_0))\right] \over 
\left[ q_0^2-(|\vec{q}|-\mu)^2-
({\Delta_1^A}(q_0)-{\Delta_1^S}(q_0))^2 \right] \left[ q_0^2-(|\vec{q}|-\mu)^2-
4({\Delta_1^A}(q_0)+2{\Delta_1^S}(q_0))^2 \right]}\nonumber\\ 
b_+(q) &=& {{\Delta_2^S}(q_0) \over q_0^2-(|\vec{q}|+\mu)^2-
4\left[{\Delta_2^A}(q_0)+2{\Delta_2^S}(q_0)\right]^2} \nonumber\\ 
&~~& + 
{ \left[ {\Delta_2^A}(q_0)-{\Delta_2^S}(q_0) \right] 
\left[ {\Delta_2^A}(q_0)+{\Delta_2^S}(q_0) \right] 
\left[ {\Delta_2^A}(q_0)+5{\Delta_2^S}(q_0) \right] 
\over \left[q_0^2-(|\vec{q}|+\mu)^2 - 
({\Delta_2^A}(q_0)-{\Delta_2^S}(q_0))^2\right]
\left[q_0^2-(|\vec{q}|+\mu)^2- 
4({\Delta_2^A}(q_0)+2{\Delta_2^S}(q_0))^2\right] }
\nonumber\\
b_-(q) &=& {{\Delta_1^S}(q_0) \over q_0^2-(|\vec{q}|-\mu)^2-
4\left[{\Delta_1^A}(q_0)+2{\Delta_1^S}(q_0)\right]^2} \nonumber\\
&~~& + { \left[{\Delta_1^A}(q_0)-{\Delta_1^S}(q_0)\right]
\left[ {\Delta_1^A}(q_0)+{\Delta_1^S}(q_0) \right] 
\left[ {\Delta_1^A}(q_0)+5{\Delta_1^S}(q_0) \right] 
\over \left[ q_0^2-(|\vec{q}|-\mu)^2- 
({\Delta_1^A}(q_0)-{\Delta_1^S}(q_0))^2 \right]
\left[ q_0^2-(|\vec{q}|-\mu)^2-
4({\Delta_1^A}(q_0)+2{\Delta_1^S}(q_0))^2\right]}\ .
\end{eqnarray}

In a general covariant gauge, the resummed gluon propagator is 
given by
\begin{equation} \label{gluonprop}
D_{\mu\nu}(q) = {P^T_{\mu\nu}\over q^2-G(q)} + {P^L_{\mu\nu}\over q^2-F(q)} -
\xi {q_\mu q_\nu \over q^4}
\end{equation}
where $G(q)$ and $F(q)$ are functions of $q_0$ and $|\vec{q}|$ and the
projectors $P^{T,L}_{\mu\nu}$ are defined as follows:
\begin{equation} \label{PLT}
P^T_{ij} = \delta_{ij} - \hat{q_i}\hat{q_j}, \ P^T_{00}=P^T_{0i}=0, \
P^L_{\mu\nu} = -g_{\mu\nu} + {q_\mu q_\nu \over q^2} - P^T_{\mu\nu}.
\end{equation}
The functions $F$ and $G$ describe the effects of the medium
on the gluon propagator.  If we neglect the Meissner effect (that is,
if we neglect the modification
of $F(q)$ and $G(q)$ due to the gap $\Delta$ in the fermion
propagator) then $F(q)$ describes
Thomas-Fermi screening and $G(q)$ describes Landau damping and they are
given in the HDL approximation by\cite{LeBellac}
\begin{eqnarray} \label{GF}
F(q) &=& m^2 {q^2\over|\vec{q}|} \left( 1 - {iq_0\over|\vec{q}|} Q_0
\left( {iq_0\over|\vec{q}|} \right) \right), 
\ \ \ \ \ \  {\rm with}\ Q_0(x) = {1\over2} \log
\left( {x+1\over x-1} \right) \nonumber\\
G(q) &=& {1\over2} m^2 {iq_0\over|\vec{q}|} \left[ \left( 1 - \left(
{iq_0\over|\vec{q}|} \right)^2 \right) Q_0 \left( {iq_0\over|\vec{q}|}
\right) + {iq_0\over|\vec{q}|} \right] \ ,
\end{eqnarray}
where $m^2 = 3 g^2 \mu^2/2\pi^2$ is the Debye screening
mass for $N_f=3$.  We discuss the modifications of $F(q)$ and $G(q)$
due to the Meissner effect in an Appendix.

In order to obtain the final form of the gap equation,
we need 
the following trace:
%\begin{eqnarray}
\begin{equation}\label{4bigequations}
\begin{array}{l}
{\rm Tr} \left[ P_\pm(k)\gamma_\mu \left(P_+(q)a_+(q)+P_-(q)a_-(q)\right)
\gamma_\nu \right] D^{\mu\nu}(k-q) \\
\ \\
\begin{array}{lll}
& = 
a_+(q)& \left[ 2 {-1\mp\widehat{(k-q)}\cdot\hat{k}
\widehat{(k-q)}\cdot\hat{q} \over (k-q)^2-G(k-q)} +{-1\mp\hat{k}\cdot
\hat{q} {(k-q)_0^2+(\vec{k}-\vec{q})^2 \over (k-q)^2} \pm 2
\widehat{(k-q)}\cdot\hat{k} \widehat{(k-q)}\cdot\hat{q} {(k-q)_0^2 \over
(k-q)^2} \over (k-q)^2-F(k-q)} \right. \nonumber\\
&& + \left. {\xi \over (k-q)^2} \left( 1\mp\hat{k}\cdot
\hat{q} {(k-q)_0^2+(\vec{k}-\vec{q})^2 \over (k-q)^2} \pm 2
\widehat{(k-q)}\cdot\hat{k} \widehat{(k-q)}\cdot\hat{q}
{(\vec{k}-\vec{q})^2 \over (k-q)^2} \right) \right.\Biggr] \nonumber\\
\ \\ 
& + a_-(q) & \left[ 2 {-1\pm\widehat{(k-q)}\cdot\hat{k}
\widehat{(k-q)}\cdot\hat{q} \over (k-q)^2-G(k-q)} +{-1\pm\hat{k}\cdot
\hat{q} {(k-q)_0^2+(\vec{k}-\vec{q})^2 \over (k-q)^2} \mp 2
\widehat{(k-q)}\cdot\hat{k} \widehat{(k-q)}\cdot\hat{q} {(k-q)_0^2 \over
(k-q)^2} \over (k-q)^2-F(k-q)} \right. \nonumber\\
&& + \left. {\xi \over (k-q)^2} \left( 1\pm\hat{k}\cdot
\hat{q} {(k-q)_0^2+(\vec{k}-\vec{q})^2 \over (k-q)^2} \mp 2
\widehat{(k-q)}\cdot\hat{k} \widehat{(k-q)}\cdot\hat{q}
{(\vec{k}-\vec{q})^2 \over (k-q)^2} \right) \right.\Biggr]\ .
\end{array}
\end{array}
\end{equation}
%\end{eqnarray}
This allows us to recast
\eqref{4gapeq} into the following form:
\begin{equation}
\begin{array}{ll}
\Delta_{1}^A(k_0) = -{i\over 6} g^2 \begin{minipage}[l][0.5in][c]{.1in}\[\int\]
\end{minipage}\frac{d^4q}{(2\pi)^4} & \!\left[ a_+(q) 
\!\left( \!2 {-1 - \widehat{(k-q)}\cdot\hat{k}
\widehat{(k-q)}\cdot\hat{q} \over (k-q)^2-G(k-q)}\!+\!{-1 - \hat{k}\cdot
\hat{q} {(k-q)_0^2+(\vec{k}-\vec{q})^2 \over (k-q)^2}+2
\widehat{(k-q)}\cdot\hat{k} \widehat{(k-q)}\cdot\hat{q} {(k-q)_0^2 \over
(k-q)^2} \over (k-q)^2-F(k-q)} \right. \right. \\ & +\!\left. {\xi \over
(k-q)^2}\!\left( \!1\!-\!\hat{k}\cdot \hat{q} {(k-q)_0^2+(\vec{k}-\vec{q})^2
\over (k-q)^2}\!+\!2 \widehat{(k-q)}\cdot\hat{k} \widehat{(k-q)}\cdot\hat{q}
{(\vec{k}-\vec{q})^2 \over (k-q)^2} \right) \right)
\\ & + a_-(q)\!\!\left( \!2 {-1+\widehat{(k-q)}\cdot\hat{k}
\widehat{(k-q)}\cdot\hat{q} \over (k-q)^2-G(k-q)}\!+\!{-1+\hat{k}\cdot
\hat{q} {(k-q)_0^2+(\vec{k}-\vec{q})^2 \over (k-q)^2} - 2
\widehat{(k-q)}\cdot\hat{k} \widehat{(k-q)}\cdot\hat{q} {(k-q)_0^2 \over
(k-q)^2} \over (k-q)^2-F(k-q)} \right. \\ & +\!\left. \left. {\xi \over
(k-q)^2}\!\left( \!1\!+\!\hat{k}\cdot \hat{q} {(k-q)_0^2+(\vec{k}-\vec{q})^2
\over (k-q)^2}\!-\!2 \widehat{(k-q)}\cdot\hat{k} \widehat{(k-q)}\cdot\hat{q}
{(\vec{k}-\vec{q})^2 \over (k-q)^2} \right) \right) \right] \\ & \\
\Delta_{2}^A(k_0) = -{i\over 6} g^2 \begin{minipage}[l][0.5in][c]{.1in}\[\int\]
\end{minipage} \frac{d^4q}{(2\pi)^4} & \!\left[ a_+(q)
\!\left( \!2 {-1 + \widehat{(k-q)}\cdot\hat{k}
\widehat{(k-q)}\cdot\hat{q} \over (k-q)^2-G(k-q)}\!+\!{-1 + \hat{k}\cdot
\hat{q} {(k-q)_0^2+(\vec{k}-\vec{q})^2 \over (k-q)^2} - 2
\widehat{(k-q)}\cdot\hat{k} \widehat{(k-q)}\cdot\hat{q} {(k-q)_0^2 \over
(k-q)^2} \over (k-q)^2-F(k-q)} \right. \right. \\ & +\!\left. {\xi \over
(k-q)^2}\!\left( \!1\!+\!\hat{k}\cdot \hat{q} {(k-q)_0^2+(\vec{k}-\vec{q})^2
\over (k-q)^2}\!-\!2 \widehat{(k-q)}\cdot\hat{k} \widehat{(k-q)}\cdot\hat{q}
{(\vec{k}-\vec{q})^2 \over (k-q)^2} \right) \right) \\ & + a_-(q)\!\!
\left( 2 {-1 - \widehat{(k-q)}\cdot\hat{k}
\widehat{(k-q)}\cdot\hat{q} \over (k-q)^2-G(k-q)}\!+\!{-1 - \hat{k}\cdot
\hat{q} {(k-q)_0^2+(\vec{k}-\vec{q})^2 \over (k-q)^2} + 2
\widehat{(k-q)}\cdot\hat{k} \widehat{(k-q)}\cdot\hat{q} {(k-q)_0^2 \over
(k-q)^2} \over (k-q)^2-F(k-q)} \right. \\ & +\!\left. \left. {\xi \over
(k-q)^2}\!\left( \!1\!-\!\hat{k}\cdot \hat{q} {(k-q)_0^2+(\vec{k}-\vec{q})^2
\over (k-q)^2}\!+\!2 \widehat{(k-q)}\cdot\hat{k} \widehat{(k-q)}\cdot\hat{q}
{(\vec{k}-\vec{q})^2 \over (k-q)^2} \right) \right) \right]
\end{array}
\end{equation}
\addtocounter{equation}{-1}
\begin{equation} \label{GapEq}
\begin{array}{ll}
\Delta_{1}^S(k_0) = -{i\over 6} g^2 \begin{minipage}[l][0.5in][c]{.1in}\[\int\]
\end{minipage}\frac{d^4q}{(2\pi)^4} & \!\left[ b_+(q) 
\!\left( \!2 {-1 - \widehat{(k-q)}\cdot\hat{k}
\widehat{(k-q)}\cdot\hat{q} \over (k-q)^2-G(k-q)}\!+\!{-1 - \hat{k}\cdot
\hat{q} {(k-q)_0^2+(\vec{k}-\vec{q})^2 \over (k-q)^2}+2
\widehat{(k-q)}\cdot\hat{k} \widehat{(k-q)}\cdot\hat{q} {(k-q)_0^2 \over
(k-q)^2} \over (k-q)^2-F(k-q)} \right. \right. \\ & +\!\left. {\xi \over
(k-q)^2}\!\left( \!1\!-\!\hat{k}\cdot \hat{q} {(k-q)_0^2+(\vec{k}-\vec{q})^2
\over (k-q)^2}\!+\!2 \widehat{(k-q)}\cdot\hat{k} \widehat{(k-q)}\cdot\hat{q}
{(\vec{k}-\vec{q})^2 \over (k-q)^2} \right) \right)
\\ & + b_-(q)\!\!\left( \!2 {-1+\widehat{(k-q)}\cdot\hat{k}
\widehat{(k-q)}\cdot\hat{q} \over (k-q)^2-G(k-q)}\!+\!{-1+\hat{k}\cdot
\hat{q} {(k-q)_0^2+(\vec{k}-\vec{q})^2 \over (k-q)^2} - 2
\widehat{(k-q)}\cdot\hat{k} \widehat{(k-q)}\cdot\hat{q} {(k-q)_0^2 \over
(k-q)^2} \over (k-q)^2-F(k-q)} \right. \\ & +\!\left. \left. {\xi \over
(k-q)^2}\!\left( \!1\!+\!\hat{k}\cdot \hat{q} {(k-q)_0^2+(\vec{k}-\vec{q})^2
\over (k-q)^2}\!-\!2 \widehat{(k-q)}\cdot\hat{k} \widehat{(k-q)}\cdot\hat{q}
{(\vec{k}-\vec{q})^2 \over (k-q)^2} \right) \right) \right] \\ & \\
\Delta_{2}^S(k_0) = -{i\over 6} g^2 \begin{minipage}[l][0.5in][c]{.1in}\[\int\]
\end{minipage} \frac{d^4q}{(2\pi)^4} & \!\left[ b_+(q)
\!\left( \!2 {-1 + \widehat{(k-q)}\cdot\hat{k}
\widehat{(k-q)}\cdot\hat{q} \over (k-q)^2-G(k-q)}\!+\!{-1 + \hat{k}\cdot
\hat{q} {(k-q)_0^2+(\vec{k}-\vec{q})^2 \over (k-q)^2} - 2
\widehat{(k-q)}\cdot\hat{k} \widehat{(k-q)}\cdot\hat{q} {(k-q)_0^2 \over
(k-q)^2} \over (k-q)^2-F(k-q)} \right. \right. \\ & +\!\left. {\xi \over
(k-q)^2}\!\left( \!1\!+\!\hat{k}\cdot \hat{q} {(k-q)_0^2+(\vec{k}-\vec{q})^2
\over (k-q)^2}\!-\!2 \widehat{(k-q)}\cdot\hat{k} \widehat{(k-q)}\cdot\hat{q}
{(\vec{k}-\vec{q})^2 \over (k-q)^2} \right) \right) \\ & + b_-(q)\!\!
\left( 2 {-1 - \widehat{(k-q)}\cdot\hat{k}
\widehat{(k-q)}\cdot\hat{q} \over (k-q)^2-G(k-q)}\!+\!{-1 - \hat{k}\cdot
\hat{q} {(k-q)_0^2+(\vec{k}-\vec{q})^2 \over (k-q)^2} + 2
\widehat{(k-q)}\cdot\hat{k} \widehat{(k-q)}\cdot\hat{q} {(k-q)_0^2 \over
(k-q)^2} \over (k-q)^2-F(k-q)} \right. \\ & +\!\left. \left. {\xi \over
(k-q)^2}\!\left( \!1\!-\!\hat{k}\cdot \hat{q} {(k-q)_0^2+(\vec{k}-\vec{q})^2
\over (k-q)^2}\!+\!2 \widehat{(k-q)}\cdot\hat{k} \widehat{(k-q)}\cdot\hat{q}
{(\vec{k}-\vec{q})^2 \over (k-q)^2} \right) \right) \right].
\end{array}
\end{equation}

%Hence, \eqref{GapEq} makes it explicitly clear that for a self
%consistent gap equation, we need to have both color ${\bar{\bf
%3}_A}$ and color ${\bf 6}_S$ channels present in parameterizing the gap. Also,
%just like discussed in Ref. (SW), because only $a_-(q)$ and $b_-(q)$
%have singularities on the Fermi surface, we can drop all terms
%containing $a_+(q)$ and $b_+(q)$ from \eqref{GapEq} in the weak coupling
%limit. Therefore, we end up with two gap parameters, $\Delta_{1}(k_0)$
%and $\Delta_{3}(k_0)$, which are gauge independent and thus lead to
%physical gaps. The other two gap parameters, $\Delta_{2}(k_0)$
%and $\Delta_{4}(k_0)$, have nonzero magnitude but are gauge dependent
%and do not lead to physical gaps on the Fermi surface.

\section{Solving the gap equation}

In order to obtain a tractable numerical problem, we
make two further simplifying assumptions:

\begin{itemize}
\item
First, at weak coupling we expect the physics to be dominated by 
particles and holes near the Fermi surface. This manifests itself
in Eq. (\ref{GapEq}) in the fact that $a_-$ and $b_-$
have singularities on the Fermi surface while $a_+$ and $b_+$
are regular there, and we therefore expect that at weak coupling
we can neglect $a_+$ and $b_+$.  Upon doing this, 
we have equations for $\Delta_1^{A,S}$ which do not involve
$\Delta_2^{A,S}$.  We are only interested in $\Delta_1^{A,S}$,
since $\Delta_2^{A,S}$ describe the propagation of antiparticles
far from the Fermi surface.  If we assume 
that we are at weak enough coupling that $a_+$ and $b_+$ can
be neglected (that is if we assume that $\Delta_1^{A,S}\ll \mu$)
then we can ignore $\Delta_2^{A,S}$ in our calculation of $\Delta_1^{A,S}$.
(Note that we are not assuming that 
$\Delta_2^{A,S}$ is any smaller than $\Delta_1^{A,S}$;
there is no reason for this to be true.)
We will see that our results break down for $g\gtrsim 0.8$, at which 
$\Delta<10^{-7}\mu$.  Because $\Delta\ll\mu$, neglecting
the effects of $\Delta_2^{A,S}$  on  $\Delta_1^{A,S}$ 
should be a good approximation, and we do not expect that including
these effects would cure the problems we discover.
This should, however,
be investigated further.
\item
Second, we set $\Delta_1^S=0$, and solve an equation
for $\Delta_1^A$ alone.
This assumption is in fact inconsistent, as
the gap in the symmetric channel must be nonzero. This is
clear from explicit examination of the gap 
equations \eqref{GapEq}  (and indeed 
of the gap equations of Ref. \cite{CFL}).
In fact, this result is manifest
on symmetry grounds \cite{Unlocking,PisarskiRischke4}: in the presence
of $\Delta_1^A\neq 0$, a nonzero $\Delta_1^S$ breaks no new
global symmetries and there is therefore no symmetry to keep it zero.
Because single-gluon
exchange is repulsive in the symmetric channel, this condensate can only
exist in the presence of condensation in the antisymmetric channel.
Explicit calculation \cite{CFL,Schaefer,ShovkovyWije} 
shows that the symmetric
condensates are much smaller than those in the antisymmetric channels.
We are therefore confident that keeping $\Delta_1^S$ would yield
only a very small correction to $\Delta_1^A$.
\end{itemize}

We must now solve a single gap equation for $\Delta_1^A(k_0)$,
which henceforth we denote simply as $\Delta(k_0)$.  
The reader will see below that this equation is still rather involved.
Most authors have made further
approximations, valid for $g\rightarrow 0$.  Because
we make no further approximations, our results cannot
be gauge invariant. This allows us to test the claim
that the results become gauge invariant in the limit $g\rightarrow 0$,
and to use the rapidity of the disappearance of gauge dependence
as this limit is approached to evaluate at what $g$ the contributions 
we have truncated can legitimately be ignored. 

In order to obtain numerical solutions,
it is convenient to do a Wick rotation $q_0\rightarrow i q_0$
to Euclidean space, yielding the gap equation
\begin{eqnarray} \label{Delta1eq}
%\begin{array}{ll}
\Delta(k_0) &=& {g^2\over 6}\!\int\!\frac{d^4q}{(2\pi)^4} \Biggl[
{\Delta(q_0) \over q_0^2+(|\vec{q}|-\mu)^2+ 4\Delta^2(q_0)} \nonumber\\
&~&\qquad\qquad +
{\Delta(q_0) \left( q_0^2+(|\vec{q}|-\mu)^2+5\Delta^2(q_0)\right) \over
\left( q_0^2+(|\vec{q}|-\mu)^2+\Delta^2(q_0) \right) \left(
q_0^2+(|\vec{q}|-\mu)^2+4\Delta^2(q_0) \right)} \Biggr] \nonumber\\ 
 &~& \qquad\Biggl[ 2 {1-\widehat{(k-q)}\cdot\hat{k} 
\widehat{(k-q)}\cdot\hat{q} \over
(k-q)_0^2+(\vec{k}-\vec{q})^2+G(k_0-q_0,|\vec{k}-\vec{q}|)} \nonumber\\
&~&\qquad\qquad
+{1+\hat{k}\cdot\hat{q} {-(k-q)_0^2+(\vec{k}-\vec{q})^2 \over
(k-q)_0^2+(\vec{k}-\vec{q})^2} + 2\widehat{(k-q)}\cdot\hat{k}
\widehat{(k-q)}\cdot\hat{q} {(k-q)_0^2 \over
(k-q)_0^2+(\vec{k}-\vec{q})^2} \over
(k-q)_0^2+(\vec{k}-\vec{q})^2+F(k_0-q_0,|\vec{k}-\vec{q}|)} \nonumber\\
&~&\qquad\qquad+
\xi { -1+\hat{k}\cdot
\hat{q} {-(k-q)_0^2+(\vec{k}-\vec{q})^2 \over (k-q)_0^2+(\vec{k}-\vec{q})^2} - 
2 \widehat{(k-q)}\cdot\hat{k} \widehat{(k-q)}\cdot\hat{q}
{(\vec{k}-\vec{q})^2 \over (k-q)_0^2+(\vec{k}-\vec{q})^2} \over
(k-q)_0^2+(\vec{k}-\vec{q})^2 } \Biggr].
%\end{array}
\end{eqnarray}
The integral over the azimuthal
angle $\phi$ is trivial, and we therefore 
have three integrals to do.  We do the remaining angular integral
analytically, after making a change of variables.  
We define 
\[ \vec{q'}=\vec{k}-\vec{q} \]
because the integration over the polar angle $\theta$ is simpler
when the momentum integration is done over $\vec{q'}$. The
simplification arises because there is no longer any
angular dependence in the functions $F$ and $G$:
$$
F(k_0-q_0,|\vec{k}-\vec{q}|)=F(k_0-q_0,|\vec{q'}|)
$$
and similarly for $G$. 
After doing the angular integral, the gap equation reduces
to a double integral equation with integration variables
$|\vec{q'}|$ (which we henceforth denote $q$) and $q_0$:
%The gap equation is
\begin{eqnarray} \label{finalDeltaeq}
%\begin{array}{ll}
\Delta(k_0) &=& {g^2\over 48\pi^3} \int_{-\infty}^{\infty} dq_0
\int_0^\infty dq \left[ {\Delta(q_0) \over (k_0-q_0)^2+q^2+G(k_0-q_0,q)}
I_G(q_0,q) \right. \nonumber\\
 &~&\qquad \left. + {\Delta(q_0) \over (k_0-q_0)^2+q^2+F(k_0-q_0,q)}
I_F(k_0,q_0,q) + \xi {q \Delta(q_0) \over (k_0-q_0)^2+q^2}
I_\xi(k_0,q_0,q) \right]\\
{\rm where}\nonumber
%\end{array}
\end{eqnarray}
%where
\begin{eqnarray}
I_G(q_0,q < \mu) &=& {2 (q_0^2+4\Delta^2(q_0)+q^2)
(q^2+4\mu^2-q_0^2-4\Delta^2(q_0)) \over 3 q \mu^2 \sqrt{q_0^2+4\Delta^2(q_0)}}
\arctan{q \over \sqrt{q_0^2+4\Delta^2(q_0)}} \nonumber\\ &+& {4
(q_0^2+\Delta^2(q_0)+q^2) (q^2+4\mu^2-q_0^2-\Delta^2(q_0)) \over 3 q
\mu^2 \sqrt{q_0^2+\Delta^2(q_0)}} \arctan{q \over
\sqrt{q_0^2+\Delta^2(q_0)}} \nonumber\\ &+& 
{12 \Delta^2(q_0) + 6 q_0^2 - 2 q^2 - 24
\mu^2 \over 3 \mu^2} \nonumber
\end{eqnarray}

\begin{eqnarray}
&~& I_F(k_0,q_0,q < \mu) =\nonumber\\
&~&\qquad {2 ((q_0^2+4\Delta^2(q_0)) (k_0-q_0)^2-q^4)
(q^2-4\mu^2 + q_0^2+4\Delta^2(q_0)) \over 3 q \mu^2 \sqrt{q_0^2+4\Delta^2(q_0)}
((k_0-q_0)^2+q^2)} \arctan{q \over \sqrt{q_0^2+4\Delta^2(q_0)}}
\nonumber\\ &~&\qquad + 
{4 ((q_0^2+\Delta^2(q_0)) (k_0-q_0)^2-q^4) (q^2-4\mu^2 +
q_0^2+\Delta^2(q_0)) \over 3 q \mu^2 \sqrt{q_0^2+\Delta^2(q_0)}
((k_0-q_0)^2+q^2)} \arctan{q \over \sqrt{q_0^2+\Delta^2(q_0)}} \nonumber\\ 
&~&\qquad +
{6 q^4+2 (k_0-q_0)^2 (-2 q^2+12\mu^2-3q_0^2-6\Delta^2(q_0)) \over 3
\mu^2 ((k_0-q_0)^2+q^2)}  \nonumber
\end{eqnarray}

\begin{eqnarray}
&~& I_\xi(k_0,q_0,q < \mu) =\nonumber\\ 
&~&\qquad -{2 (q_0^2+4\Delta^2(q_0)-(k_0-q_0)^2) (q^2-4\mu^2 +
q_0^2+4\Delta^2(q_0)) \over 3 \mu^2 \sqrt{q_0^2+4\Delta^2(q_0)}
((k_0-q_0)^2+q^2)} \arctan{q \over \sqrt{q_0^2+4\Delta^2(q_0)}} 
\nonumber\\ &~&\qquad - 
{4 (q_0^2+\Delta^2(q_0)-(k_0-q_0)^2) (q^2-4\mu^2 +
q_0^2+\Delta^2(q_0)) \over 3 \mu^2 \sqrt{q_0^2+\Delta^2(q_0)}
((k_0-q_0)^2+q^2)} \arctan{q \over \sqrt{q_0^2+\Delta^2(q_0)}} \nonumber\\ 
&~&\qquad +
{2 q (2 q^2-3(k_0-q_0)^2-12\mu^2+3q_0^2+6\Delta^2(q_0)) \over 3 \mu^2
((k_0-q_0)^2+q^2)} \nonumber
\end{eqnarray}

\begin{eqnarray}
&~&I_G(q_0,q \ge\mu) =\nonumber\\
&~&\quad{(q_0^2+4\Delta^2(q_0)+q^2)
(q^2+4\mu^2-q_0^2-4\Delta^2(q_0)) \over 3 q \mu^2
\sqrt{q_0^2+4\Delta^2(q_0)}} \left( \arctan{q \over 
\sqrt{q_0^2+4\Delta^2(q_0)}}\right.\nonumber\\&~&
\qquad\qquad\qquad\qquad\qquad\qquad\qquad\qquad\qquad\qquad\qquad\left.
-\arctan{q - 2\mu \over
\sqrt{q_0^2+4\Delta^2(q_0)}} \right) \nonumber\\ &~&\quad 
+ {2 (q_0^2+\Delta^2(q_0)+q^2)
(q^2+4\mu^2-q_0^2-\Delta^2(q_0)) \over 3 q \mu^2
\sqrt{q_0^2+\Delta^2(q_0)}} \left( \arctan{q \over 
\sqrt{q_0^2+\Delta^2(q_0)}} \right.\nonumber\\&~&
\qquad\qquad\qquad\qquad\qquad\qquad\qquad\qquad\qquad\qquad\qquad\left.
- \arctan{q - 2\mu \over
\sqrt{q_0^2+\Delta^2(q_0)}} \right) \nonumber\\ &~&\quad
+ {4 (q_0^2+\Delta^2(q_0)+q^2) \over
3 q \mu} \ln{q_0^2+\Delta^2(q_0)+q^2 \over
q_0^2+\Delta^2(q_0)+(q-2\mu)^2} \nonumber\\ &~&\quad
+ {2 (q_0^2+4\Delta^2(q_0)+q^2) \over 3
q \mu} \ln{q_0^2+4\Delta^2(q_0)+q^2 \over
q_0^2+4\Delta^2(q_0)+(q-2\mu)^2} \nonumber\\ &~&\quad 
+ {12 \Delta^2(q_0) + 6 q_0^2 -
6 q^2 - 8 \mu^2 -12 \mu q \over 3 \mu q} \nonumber
\end{eqnarray}

\begin{eqnarray}
&~&I_F(k_0,q_0,q \ge \mu)=\nonumber\\ 
&~&\quad{((q_0^2+4\Delta^2(q_0))
(k_0-q_0)^2-q^4) (q^2-4\mu^2 + q_0^2+4\Delta^2(q_0)) \over 3 q \mu^2
\sqrt{q_0^2+4\Delta^2(q_0)} ((k_0-q_0)^2+q^2)} \left( \arctan{q \over
\sqrt{q_0^2+4\Delta^2(q_0)}} \right.\nonumber\\&~&
\qquad\qquad\qquad\qquad\qquad\qquad\qquad\qquad\qquad\qquad\qquad\left.
- \arctan{q - 2\mu \over
\sqrt{q_0^2+4\Delta^2(q_0)}} \right) \nonumber\\ &~&\quad + {2 
((q_0^2+\Delta^2(q_0)) (k_0-q_0)^2-q^4) (q^2-4\mu^2 +
q_0^2+\Delta^2(q_0)) \over 3 q \mu^2 \sqrt{q_0^2+\Delta^2(q_0)}
((k_0-q_0)^2+q^2)} \left( \arctan{q \over \sqrt{q_0^2+\Delta^2(q_0)}} 
\right.\nonumber\\&~&
\qquad\qquad\qquad\qquad\qquad\qquad\qquad\qquad\qquad\qquad\qquad\left.
-
\arctan{q - 2\mu \over \sqrt{q_0^2+\Delta^2(q_0)}} \right)
\nonumber\\ &~&\quad + {4 (q^4 -
(q_0^2+\Delta^2(q_0))(k_0-q_0)^2) \over 3 q \mu (q^2+(k_0-q_0)^2)}
\ln{q_0^2+\Delta^2(q_0)+q^2 \over q_0^2+\Delta^2(q_0)+(q-2\mu)^2} 
\nonumber\\ &~&\quad +{2
(q^4 - (q_0^2+4\Delta^2(q_0))(k_0-q_0)^2) \over 3 q \mu 
(q^2+(k_0-q_0)^2)} \ln{q_0^2+4\Delta^2(q_0)+q^2 \over
q_0^2+4\Delta^2(q_0)+(q-2\mu)^2} \nonumber\\ &~&\quad 
+ {6 q^4+2 (k_0-q_0)^2 (6 \mu
q+4\mu^2-3q_0^2-6\Delta^2(q_0)) \over 3 \mu q ((k_0-q_0)^2+q^2)} \nonumber
\end{eqnarray}
%\vfill\eject
\begin{eqnarray}
&~&I_\xi(k_0,q_0,q \ge \mu) =\nonumber\\  
&~&\quad-{(q_0^2+4\Delta^2(q_0)-(k_0-q_0)^2)
(q^2-4\mu^2 + q_0^2+4\Delta^2(q_0)) \over 3 \mu^2
\sqrt{q_0^2+4\Delta^2(q_0)} ((k_0-q_0)^2+q^2)} \left( \arctan{q \over
\sqrt{q_0^2+4\Delta^2(q_0)}} \right.\nonumber\\&~&
\qquad\qquad\qquad\qquad\qquad\qquad\qquad\qquad\qquad\qquad\qquad\left.
- \arctan{q - 2\mu \over
\sqrt{q_0^2+4\Delta^2(q_0)}} \right) \nonumber\\ &~&\quad - {2
(q_0^2+\Delta^2(q_0)-(k_0-q_0)^2) (q^2-4\mu^2 +  q_0^2+\Delta^2(q_0))
\over 3 \mu^2 \sqrt{q_0^2+\Delta^2(q_0)} ((k_0-q_0)^2+q^2)} \left(
\arctan{q \over \sqrt{q_0^2+\Delta^2(q_0)}} \right.\nonumber\\&~&
\qquad\qquad\qquad\qquad\qquad\qquad\qquad\qquad\qquad\qquad\qquad\left.
- \arctan{q - 2\mu \over
\sqrt{q_0^2+\Delta^2(q_0)}} \right) \nonumber\\ &~&\quad + {4
(q_0^2+\Delta^2(q_0)-(k_0-q_0)^2) \over 3 \mu (q^2+(k_0-q_0)^2)} 
\ln{q_0^2+\Delta^2(q_0)+q^2 \over q_0^2+\Delta^2(q_0)+(q-2\mu)^2}
\nonumber\\ &~&\quad + {2
(q_0^2+4\Delta^2(q_0)-(k_0-q_0)^2) \over 3 \mu (q^2+(k_0-q_0)^2)}
\ln{q_0^2+4\Delta^2(q_0)+q^2 \over q_0^2+4\Delta^2(q_0)+(q-2\mu)^2} 
\nonumber\\ &~&\quad + 
{6q_0^2+12\Delta^2(q_0)-6(k_0-q_0)^2-8\mu^2-12\mu q) \over 3 \mu
((k_0-q_0)^2+q^2)}\nonumber
\end{eqnarray}

We have solved the gap equation (\ref{finalDeltaeq}) numerically
for several different values of $g$ and several different values
of $\xi$.  
It is convenient to change integration variables from $q_0$ to
$\ln q_0$ and from $q$ to $\ln q$.  
We evaluate the $q$ integral over a 
range $q_{\rm min}<q<10^4\mu$ with $q_{\rm min}/\mu$ chosen differently
for each $g$ in such a way that it is less than $10^{-5}\Delta(0)$ in
all cases. The $q_0$ integral is made even in $q_0$ (by taking the
average of the integrand at $q_0$ and $-q_0$) and then evaluated
over a range $q_{0\rm min}<q_0<100\mu$, where we chose 
$q_{0\rm min}=q_{\rm min}$.  We have checked that
our results are insensitive to the 
choice of upper and lower cutoffs of the integration region.
It was probably not necessary to choose $q_{\rm min}$
and $q_{0\rm min}$ quite as small as we did.
It is, however, quite important to extend the upper limit of the  
$q_0$ and $q$ integrals to well above $\mu$ in order to avoid 
sensitivity to the ultraviolet cutoff.\footnote{The one exception,
in which we do find some sensitivity to one of our limits
of integration, 
is at $g=3.5576$. With $g$ this large, we should perhaps
have extended the upper cutoff of the 
$q_0$ integration to 1000 $\mu$,
as the results shown in Fig. 1 below make clear.}
We use an iterative method, in which an initial
guess for $\Delta(k_0)$ is used on the right-hand side 
of (\ref{finalDeltaeq}), the integrals are done
yielding a new $\Delta(k_0)$, which
is in turn used on the right-hand side.  The solution converges
well after about ten iterations. All results we show
were iterated at least fifteen times.  

\begin{figure}[t]
\centering
\vspace{-0.4in}
\epsfig{file=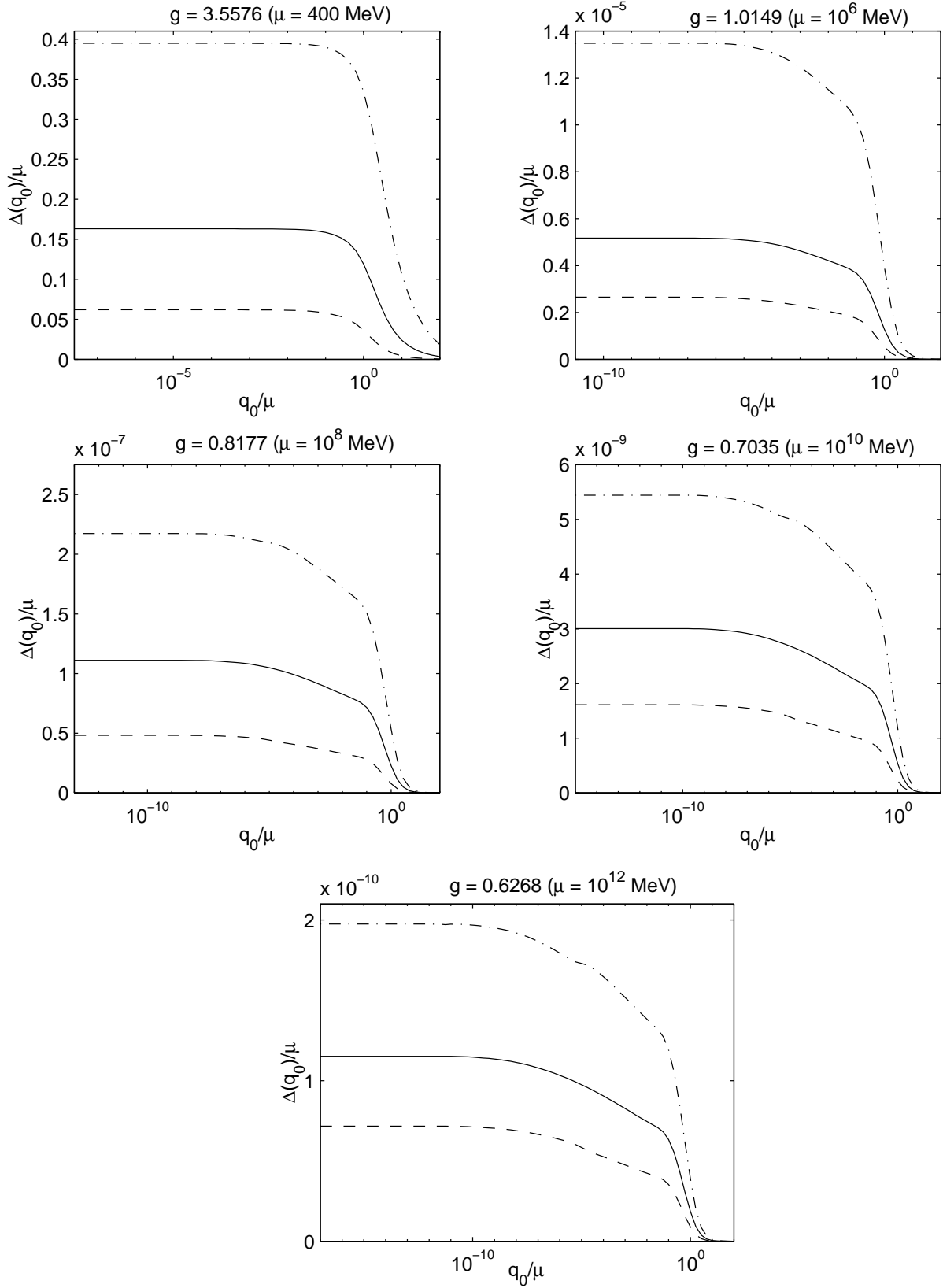,width=6.3in}%,bbllx=0,bblly=0,bburx=400,bbury=171}
\caption{The gap $\Delta(q_0)$ for five different values of the
coupling constant $g$. In each plot, the upper, middle, and lower
curves are calculations done using three different 
gauges $\xi=-1,0,1$. In each panel, the range
over which the $q_0$ integral was done is that shown.} \label{fig:3fgaps} 
\end{figure}

Our results are shown in \figref{fig:3fgaps}.  Note that
the output of our calculation is a plot of $\Delta(q_0)/\mu$
as a function of $q_0/\mu$ for some choice of $g$ and $\xi$.
The only way in which 
$\mu$ enters the calculation is to set the units of energy.
The values of 
$\mu$ shown in Fig. \ref{fig:3fgaps} corresponding to each
value of $g$ do not come from the calculation. They
are obtained by assuming that the running
coupling $g$ should be evaluated 
at the scale $\mu$ and using the one-loop beta function with 
$\Lambda_{\rm QCD}=200$ MeV.  We include these values
of $\mu$  to make comparison
with the results of Refs. \cite{SW,EHHS} easier.   If, as seems quite
reasonable, $g$ should in fact be evaluated at a $g$-dependent
scale which is lower than $\mu$, then the values of $g$
at which we have done our calculations correspond to larger
values of $\mu$ than shown in Fig. \ref{fig:3fgaps} \cite{BBSunpub}.
Evans, Hormuzdiar, Hsu, and Schwetz have obtained numerical
solutions to simplified gap equations describing the gap in
the CFL phase \cite{EHHS}.  Their results agree reasonably well with
the results of our calculation done in $\xi=0$ gauge but 
disagree qualitatively with
ours in any other gauge.
Simply setting $\xi=0$, as in Ref. \cite{SW,EHHS}, is not
a valid approximation at the values of $g$ at which we (and
these authors) work.

\begin{figure}[t] \centering
\epsfig{file=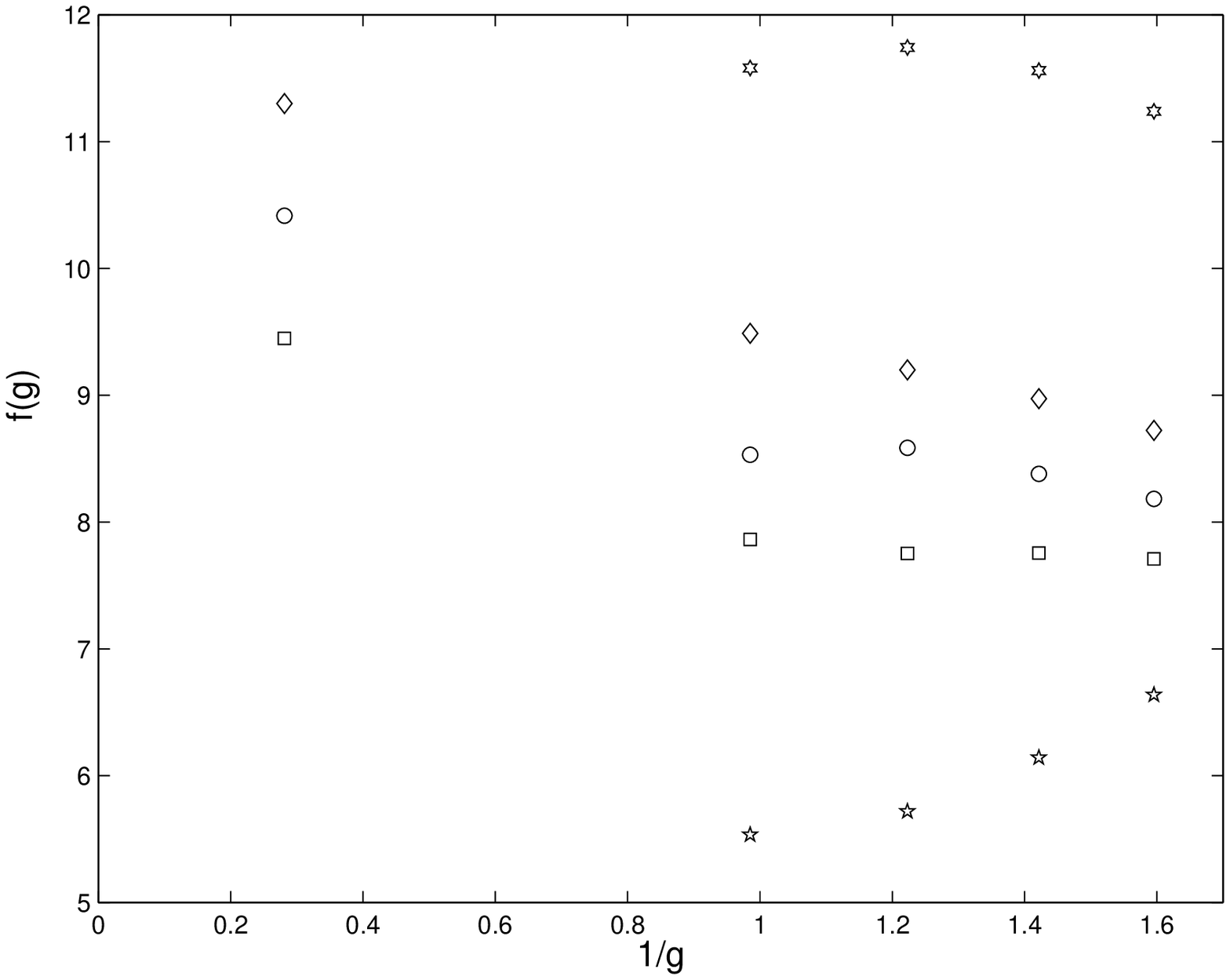,width=5.in}%,bbllx=0,bblly=0,bburx=400,bbury=171}
\caption{The function $f(g)$, defined in \eqref{fdefn}, for five
different values of the coupling constant $g$.  At each $g$, the points
(from top to bottom) correspond to different gauges with $\xi=-4,-1,0,1,4$
respectively. Note that the horizontal axis is $1/g$ and $\mu$
increases to the right. At the largest value of g, we only show
$\xi=-1,0,1$. In Fig. \ref{fig:3fgaps}, we have not shown the
$\Delta(q_0)$ curves for $\xi=\pm 4$ because in these gauges
$\Delta(q_0)$ is very small or large on the scales of
Fig. \ref{fig:3fgaps}.} \label{fig:fgxi}
\end{figure}

How should one interpret the results of a gauge dependent calculation,
given that at any fixed $g$ one can obtain any result one likes
if one is willing to explore gauge parameters $-\infty<\xi<\infty$?
In the present circumstance, the idea is that we expect this
calculation to give a gauge invariant result in the $g\rightarrow 0$
limit.  More precisely, if we define
\begin{equation}
f(g)\equiv \ln\left[\frac{\Delta(0)}{\mu}\right]+\frac{3\pi^2}{\sqrt{2}g}+
5\ln g
\label{fdefn}
\end{equation}
then we expect $f$ to go to a $\xi$-independent constant in the 
$g\rightarrow 0$ limit.  
In Fig. \ref{fig:fgxi}, we plot $f(g)$ in five different gauges.
From this figure we learn:
\begin{itemize}
\item
For any $\xi$, $f(g)$ is a reasonably slowly varying function of $g$.
This confirms Son's result (\ref{sonequation}) and justifies
an analysis in terms of $f(g)$.
\item
It does appear that $\lim_{g\rightarrow 0}f(g)$ is a $\xi$-independent
constant,
perhaps not far from the estimate of Ref. \cite{SW},
namely $\lim_{g\rightarrow 0}f(g)=8.88$, or that of Ref. \cite{rockefeller},
namely $\lim_{g\rightarrow 0}f(g)=7.84$.  
\item
If we do a calculation in some fixed gauge, we expect 
that at small enough $g$ this
calculation yields a good estimate of the true gauge invariant
result.  By doing calculations in several gauges, we can bound the
regime of applicability of this estimate. 
We can only trust our calculation of $f(g)$ in
the regime in which the $\xi$-dependence of $f$ decreases with
decreasing $g$.  
Our calculation of $f(g)$ is completely meaningless
unless $g$ is small enough that the curves for different values of $\xi$
are converging.  Fig.~\ref{fig:fgxi} shows that the gauge dependence of 
our result for $f$ is about the same for all $g\gtrsim 0.8$.
It is only for $g\lesssim 0.8$ that $f(g)$ calculated in different
gauges {\it begins} to converge.  
At larger values of $g$ our calculation provides no guide
whatsoever as to the value of $f$ that would be obtained
in a complete, gauge invariant calculation including all
the physics neglected in the present calculation. Even at
$g=0.8$ the values of $\Delta(0)$ differ by a factor of about 400 for
gaps with $\xi=-4$ and $\xi=4$. We could make the gauge dependence look
even larger by choosing larger values of $|\xi|$. Our result does not
guarantee that the calculation is under control for $g<0.8$, but it does
guarantee that the result is uncontrolled and completely meaningless for
$g>0.8$.
\end{itemize}

\section{Conclusion}

We have detailed our assumptions and approximations as we made them.
Let us now ask which of them should be improved upon if we wish
to include those contributions whose neglect we have diagnosed
via the gauge dependence of our results.  
Note that $g=0.8$ 
corresponds to $\Delta/\mu\sim 10^{-7}$.  Thus, those contributions 
to $f$ which
we have neglected which are controlled when $\Delta\ll \mu$ are {\it not}
responsible for the breakdown of our calculation around $g\sim 0.8$. 
We believe that the assumptions we made in writing the
ansatz (\ref{delta}) and the assumptions we made in neglecting
$\Delta_1^S$ and $\Delta^{A,S}_2$ all introduce errors which are small
when $\Delta\ll\mu$.  (For example, even though neglecting $\Delta_2$
is a source of gauge dependence, we do not expect that remedying
this neglect would change $f(g)$ appreciably in any gauge at 
$g\sim 0.8$, where $\Delta/\mu$ is so small.)      Hence, we believe
that it is the assumptions made in writing the truncated gap equation
(\ref{SDeq}) that are at fault.  One obvious possible explanation
is the absence of vertex corrections, although there are
other missing skeleton diagrams which should
also be investigated.  

The gap $\Delta$ is of course a gauge invariant observable.
A complete calculation would yield a gauge invariant
expression for the function $f$, which could be expanded
as a power series in $g$.  We learn
three things from our (incomplete and gauge dependent) 
calculation.
First, our results obtained
in different gauges appear to converge at small 
$g$ and support previous estimates of
$\lim_{g\rightarrow 0} f(g)$, namely the $g^0$ term in the expansion
of $f$.  Second, because the results we obtain in different
gauges only begin to converge for $g<g_c\sim 0.8$,
we learn that contributions to our gauge dependent function
$f$ which are of order $g^1$ and higher must have gauge dependent
parts which are numerically large at $g\sim g_c$. Although
we have simply evaluated $f(g)$ and not expanded it in $g$,
we learn that such an expansion is uncontrolled for $g>g_c$.
This suggests that if we knew the complete, gauge invariant
function $f$, the $g^1$ and higher terms in that expansion
would also become uncontrolled for $g>g_c$. 
It may be that the vertex corrections are the
dominant contribution to the
missing physics which is responsible for this breakdown:
this hypothesis is supported by the arguments of Ref. \cite{rockefeller}
that these effects contribute to $f$ at order $g^1$.
Regardless of whether the vertex corrections turn out to
be the most important effect left out of the truncated gap 
equation (\ref{SDeq}),  our calculation demonstrates
that some contribution which is formally subleading
is in fact large enough to render the calculation uncontrolled
at $g\sim g_c$.  The third thing we
learn is that although present calculations
do yield reasonable estimates of $\lim_{g\rightarrow 0} f(g)$,
if one is interested in using these calculations to estimate the value
of $\Delta$ to within a factor of two, this can only be
done for $g\ll g_c\sim 0.8$.

In the CFL phase, all eight gluons get a mass.  This means that in the
CFL phase there are no gapless fermionic excitations, and
no massless gluonic excitations, and therefore
no non-Abelian physics in the infrared  to obstruct
weak-coupling calculations.   
The lesson we have learned is that even though everything
is in principle under control,
present weak-coupling calculations break
down for $g>g_c\sim 0.8$, corresponding to
$\mu<\mu_c$ with $\mu_c\sim 10^8$ MeV (or higher \cite{BBSunpub}). 
This break down occurs even though $\Delta\ll\mu$ at $g\sim g_c$.
It should be noted that what breaks down is the 
weak-coupling calculation of the magnitude of 
the gap $\Delta$.  Estimates based
on models normalized to give reasonable zero density phenomenology
can still be used as a guide, albeit a qualitative one.
Furthermore, regardless of the fact that a 
controlled calculation of $\Delta$ has not
yet been done at $\mu<10^8$ MeV,
it is possible to construct a controlled effective field
theory which describes the infrared 
physics of the CFL phase on length scales long compared to $1/\Delta$,
since in such an effective theory $\Delta$ is simply a parameter
determined by physics outside the effective theory.
This infrared physics is dominated by the massless Abelian
gauge bosons \cite{CFL,MagFields},
the Nambu-Goldstone boson arising from spontaneously broken
$U(1)_B$ \cite{CFL}, and the pseudo-Nambu-Goldstone bosons
arising from spontaneously broken chiral symmetry  
which have small masses due to the nonvanishing quark
masses \cite{CFL,Casalbuoni,SonStephanov,RWZ,HLM,MT,RSWZ,Zarembo,BBS}.

\acknowledgments

We thank I. Shovkovy for suggesting that gauge dependence
could be used as a diagnostic device and thank
T. Schaefer for very helpful discussions.
We are grateful to the Department of
Energy's Institute for Nuclear Theory
at the University of Washington for generous hospitality and
support during the completion of this work.
This research is also supported in part  by the Department
of Energy under cooperative research agreement DF-FC02-94ER40818.
The work of KR is supported in part by
by a DOE OJI grant and by the Alfred P. Sloan Foundation.

\appendix

\section{The Meissner Effect}

In this appendix, we set up the calculation of the Meissner
effect. That is, we investigate the effect of the presence of a gap $\Delta$
on the functions $F$ and $G$ which describe the screening
of the gluon propagator.  

In order to establish some necessary notation, 
we must begin by filling in some details in the derivation
of Eq. (\ref{4gapeq}) from Eq. (\ref{SDeq}).
We 
work in a
color-flavor basis $(\{i,a\},\{j,b\})$. In this basis, we define the
following two $9 \times 9$ matrices:
\begin{equation} \label{P}
Q^{ab}_{ij} = (\lambda^A_I)^{ab}(\lambda^A_I)_{ij} = 
\left(\begin{array}{ccccccccc}
 0 & 1 & 1 & & & & & & \\
1 &  0 & 1 & & & & & & \\
1 & 1 &  0 & & & & & & \\
 & & & 0 & -1 & & & & \\
 & & & -1 & 0 & & & & \\
 & & & & & 0 & -1 & & \\
 & & & & & -1 & 0 & & \\
 & & & & & & & 0 & -1 \\
 & & & & & & & -1 & 0 
\end{array}\right)
\end{equation}
\begin{equation} \label{Q}
R^{ab}_{ij} = (\lambda^S_J)^{ab}(\lambda^S_J)_{ij} =
\left(\begin{array}{ccccccccc}
 2 & 1 & 1 & & & & & & \\
 1 & 2 & 1 & & & & & & \\
 1 & 1 & 2 & & & & & & \\
 & & & 0 & 1 & & & & \\
 & & & 1 & 0 & & & & \\
 & & & & & 0 & 1 & & \\
 & & & & & 1 & 0 & & \\
 & & & & & & & 0 & 1 \\
 & & & & & & & 1 & 0 
\end{array}\right)
\end{equation}
which represent the antisymmetric color and flavor ${\bar{\bf 3}}_A$ 
and the symmetric color and flavor ${\bf 6}_S$ channels
respectively in this basis. 

In the derivation of the gap equation, we were only interested
in the off-diagonal lower left component of the Nambu-Gorkov
fermion propagator $S$.
However, the calculation of the Meissner effect involves
all components of the fermion propagator. 
Obtaining
the fermion propagator by inverting the inverse propagator (\ref{Sinv}) is
straightforward but tedious. After a lot of algebra and using the ansatz 
(\ref{delta}) for the gap matrix, we find:
\begin{eqnarray}\label{entireS}
S(q)=\left(\begin{array}{cc}
S_{11}(q) & S_{12}(q)\\
S_{21}(q) & S_{22}(q)
\end{array}\right)
\end{eqnarray}
where
\begin{eqnarray} \label{S11}
S_{11}(q) = \left(\begin{array}{ccccccccc}
 A(q) & B(q) & B(q) & & & & & & \\
 B(q) & A(q) & B(q) & & & & & & \\
 B(q) & B(q) & A(q) & & & & & & \\
 & & & C(q) & & & & & \\
 & & & & C(q) & & & & \\
 & & & & & C(q) & & & \\
 & & & & & & C(q) & & \\
 & & & & & & & C(q) & \\
 & & & & & & & & C(q)
\end{array}\right) \\ \label{S22}
S_{22}(q) = \left(\begin{array}{ccccccccc}
 E(q) & H(q) & H(q) & & & & & & \\
 H(q) & E(q) & H(q) & & & & & & \\
 H(q) & H(q) & E(q) & & & & & & \\
 & & & D(q) & & & & & \\
 & & & & D(q) & & & & \\
 & & & & & D(q) & & & \\
 & & & & & & D(q) & & \\
 & & & & & & & D(q) & \\
 & & & & & & & & D(q)
\end{array}\right) \\ \label{S21}
S_{21}(q) = \overline{S_{12}}(q) = -\left(\begin{array}{ccccccccc}
 K(q) & L(q) & L(q) & & & & & & \\
 L(q) & K(q) & L(q) & & & & & & \\
 L(q) & L(q) & K(q) & & & & & & \\
 & & & 0 & M(q) & & & & \\
 & & & M(q) & 0 & & & & \\
 & & & & & 0 & M(q) & & \\
 & & & & & M(q) & 0 & & \\
 & & & & & & & 0 & M(q) \\
 & & & & & & & M(q) & 0
\end{array}\right)
\end{eqnarray}
and where the above functions are defined as follows:
\begin{equation} 
\begin{array}{ll}
A(q) &= \gamma^0 \left[ P_+(q) {q_0-\mu-|\vec{q}|\over
q_0^2-(|\vec{q}|+\mu)^2-4({\Delta_2^A}(q_0)+2{\Delta_2^S}(q_0))^2}
{q_0^2-(|\vec{q}|+\mu)^2-3\left({\Delta_2^A}(q_0)\right)^2
-11\left({\Delta_2^S}(q_0)\right)^2
-10{\Delta_2^A}(q_0){\Delta_2^S}(q_0) \over
q_0^2-(|\vec{q}|+\mu)^2-({\Delta_2^A}(q_0)-{\Delta_2^S}(q_0))^2}
\right. \\ &+ \left. P_-(q) {q_0-\mu+|\vec{q}|\over
q_0^2-(|\vec{q}|-\mu)^2-4({\Delta_1^A}(q_0)+2{\Delta_1^S}(q_0))^2}
{q_0^2-(|\vec{q}|-\mu)^2-3\left({\Delta_1^A}(q_0)\right)^2
-11\left({\Delta_1^S}(q_0)\right)^2
-10{\Delta_1^A}(q_0){\Delta_1^S}(q_0) \over
q_0^2-(|\vec{q}|-\mu)^2-({\Delta_1^A}(q_0)-{\Delta_1^S}(q_0))^2} 
\right] \\  & \\ 
B(q) &= \gamma^0 \left[ P_+(q) {q_0-\mu-|\vec{q}|\over
q_0^2-(|\vec{q}|+\mu)^2-4({\Delta_2^A}(q_0)+2{\Delta_2^S}(q_0))^2}
{\left({\Delta_2^A}(q_0)+5{\Delta_2^S}(q_0)\right)
\left({\Delta_2^A}(q_0)+{\Delta_2^S}(q_0)\right) 
\over q_0^2-(|\vec{q}|+\mu)^2-({\Delta_2^A}(q_0)-{\Delta_2^S}(q_0))^2} 
\right. \\ &+ \left. P_-(q) {q_0-\mu+|\vec{q}|\over
q_0^2-(|\vec{q}|-\mu)^2-4({\Delta_1^A}(q_0)+2{\Delta_1^S}(q_0))^2} 
{\left({\Delta_1^A}(q_0)+5{\Delta_1^S}(q_0)\right)
\left({\Delta_1^A}(q_0)+{\Delta_1^S}(q_0)\right) 
\over q_0^2-(|\vec{q}|-\mu)^2-({\Delta_1^A}(q_0)-{\Delta_1^S}(q_0))^2} 
\right] %\\ & \\ 
\end{array}
\end{equation}
\addtocounter{equation}{-1}
\begin{equation} \label{AK}
\begin{array}{ll}
C(q) &= \gamma^0 \left[ P_+(q) {q_0-\mu-|\vec{q}|\over
q_0^2-(|\vec{q}|+\mu)^2-({\Delta_2^A}(q_0)-{\Delta_2^S}(q_0))^2} + P_-(q)
{q_0-\mu+|\vec{q}|\over
q_0^2-(|\vec{q}|-\mu)^2-({\Delta_1^A}(q_0)-{\Delta_1^S}(q_0))^2} 
\right] \\  & \\ 
D(q) &= C\gamma^0 \left[ P_-(q) {q_0+\mu+|\vec{q}|\over
q_0^2-(|\vec{q}|+\mu)^2-({\Delta_2^A}(q_0)-{\Delta_2^S}(q_0))^2} +
P_+(q) {q_0+\mu-|\vec{q}|\over
q_0^2-(|\vec{q}|-\mu)^2-({\Delta_1^A}(q_0)-{\Delta_1^S}(q_0))^2}\right]
C \\  & \\
E(q) &= C\gamma^0 \left[ P_-(q) {q_0+\mu+|\vec{q}|\over
q_0^2-(|\vec{q}|+\mu)^2-4({\Delta_2^A}(q_0)+2{\Delta_2^S}(q_0))^2}
{q_0^2-(|\vec{q}|+\mu)^2-3\left({\Delta_2^A}(q_0)\right)^2
-11\left({\Delta_2^S}(q_0)\right)^2 -10{\Delta_2^A}(q_0){\Delta_2^S}(q_0) \over
q_0^2-(|\vec{q}|+\mu)^2-({\Delta_2^A}(q_0)-{\Delta_2^S}(q_0))^2}
\right. \\ &+ \left. P_+(q) {q_0+\mu-|\vec{q}|\over
q_0^2-(|\vec{q}|-\mu)^2-4({\Delta_1^A}(q_0)+2{\Delta_1^S}(q_0))^2} 
{q_0^2-(|\vec{q}|-\mu)^2-3\left({\Delta_1^A}(q_0)\right)^2
-11\left({\Delta_1^S}(q_0)\right)^2
-10{\Delta_1^A}(q_0){\Delta_1^S}(q_0) \over
q_0^2-(|\vec{q}|-\mu)^2-({\Delta_1^A}(q_0)-{\Delta_1^S}(q_0))^2} 
\right] C \\  & \\ 
H(q) &= C\gamma^0 \left[ P_-(q) {q_0+\mu+|\vec{q}|\over
q_0^2-(|\vec{q}|+\mu)^2-4({\Delta_2^A}(q_0)+2{\Delta_2^S}(q_0))^2}
{\left({\Delta_2^A}(q_0)+5{\Delta_2^S}(q_0)\right)
\left({\Delta_2^A}(q_0)+{\Delta_2^S}(q_0)\right) 
\over q_0^2-(|\vec{q}|+\mu)^2-({\Delta_2^A}(q_0)-{\Delta_2^S}(q_0))^2} 
\right. \\ &+ \left. P_+(q) {q_0+\mu-|\vec{q}|\over
q_0^2-(|\vec{q}|-\mu)^2-4({\Delta_1^A}(q_0)+2{\Delta_1^S}(q_0))^2} 
{\left({\Delta_1^A}(q_0)+5{\Delta_1^S}(q_0)\right)
\left({\Delta_1^A}(q_0)+{\Delta_1^S}(q_0)\right) \over
q_0^2-(|\vec{q}|-\mu)^2-({\Delta_1^A}(q_0)-{\Delta_1^S}(q_0))^2}\right] C
\\ & \\
K(q) &= 2C\gamma^5 \left[ P_+(q) \left({{\Delta_2^S}(q_0)\over
q_0^2-(|\vec{q}|+\mu)^2-4({\Delta_2^A}(q_0)+2{\Delta_2^S}(q_0))^2}
\right.\right. \\ & \qquad\qquad\qquad \left. +
{{\Delta_2^A}(q_0)-{\Delta_2^S}(q_0) \over
q_0^2-(|\vec{q}|+\mu)^2-4({\Delta_2^A}(q_0)+2{\Delta_2^S}(q_0))^2}
{\left({\Delta_2^A}(q_0)+5{\Delta_2^S}(q_0)\right)
\left({\Delta_2^A}(q_0)+{\Delta_2^S}(q_0)\right) \over
q_0^2-(|\vec{q}|+\mu)^2-({\Delta_2^A}(q_0)-{\Delta_2^S}(q_0))^2} \right)
\\ &+  \left.  P_-(q) \left({{\Delta_1^S}(q_0)\over
q_0^2-(|\vec{q}|-\mu)^2-4({\Delta_1^A}(q_0)-2{\Delta_1^S}(q_0))^2} 
\right.\right. \\ & \qquad\qquad\qquad \left.\left. +
{{\Delta_1^A}(q_0)-{\Delta_1^S}(q_0) \over 
q_0^2-(|\vec{q}|-\mu)^2-4({\Delta_1^A}(q_0)+2{\Delta_1^S}(q_0))^2}
{\left({\Delta_1^A}(q_0)+5{\Delta_1^S}(q_0)\right)
\left({\Delta_1^A}(q_0)+{\Delta_1^S}(q_0)\right) 
\over q_0^2-(|\vec{q}|-\mu)^2-({\Delta_1^A}(q_0)-{\Delta_1^S}(q_0))^2} \right)
\right] \\  & \\ 
L(q) &= C\gamma^5 \left[ P_+(q) \left({{\Delta_2^S}(q_0)+{\Delta_2^A}(q_0)\over
q_0^2-(|\vec{q}|+\mu)^2-4({\Delta_2^A}(q_0)+2{\Delta_2^S}(q_0))^2} 
\right.\right. \\ & \qquad\qquad\qquad \left.\left.+
{-{\Delta_2^A}(q_0)+{\Delta_2^S}(q_0) \over
q_0^2-(|\vec{q}|+\mu)^2-4({\Delta_2^A}(q_0)+2{\Delta_2^S}(q_0))^2}
{\left({\Delta_2^A}(q_0)+5{\Delta_2^S}(q_0)\right)
\left({\Delta_2^A}(q_0)+{\Delta_2^S}(q_0)\right) \over
q_0^2-(|\vec{q}|+\mu)^2-({\Delta_2^A}(q_0)-{\Delta_2^S}(q_0))^2} \right)
\right. \\ &+ \left. P_-(q) \left({{\Delta_1^S}(q_0)+{\Delta_1^A}(q_0)\over
q_0^2-(|\vec{q}|-\mu)^2-4({\Delta_1^A}(q_0)+2{\Delta_1^S}(q_0))^2} 
\right.\right. \\ & \qquad\qquad\qquad \left.\left.+
{-{\Delta_1^A}(q_0)+{\Delta_1^S}(q_0) \over 
q_0^2-(|\vec{q}|-\mu)^2-4({\Delta_1^A}(q_0)+2{\Delta_1^S}(q_0))^2}
{\left({\Delta_1^A}(q_0)+5{\Delta_1^S}(q_0)\right)
\left({\Delta_1^A}(q_0)+{\Delta_1^S}(q_0)\right) 
\over q_0^2-(|\vec{q}|-\mu)^2-({\Delta_1^A}(q_0)-{\Delta_1^S}(q_0))^2} \right)
\right] \\  & \\ 
M(q) &= C\gamma^5 \left[ P_+(q) {-{\Delta_2^A}(q_0)+{\Delta_2^S}(q_0) \over
q_0^2-(|\vec{q}|+\mu)^2-({\Delta_2^A}(q_0)-{\Delta_2^S}(q_0))^2} + P_-(q)
{-{\Delta_1^A}(q_0)+{\Delta_1^S}(q_0) \over
q_0^2-(|\vec{q}|-\mu)^2-({\Delta_1^A}(q_0)-{\Delta_1^S}(q_0))^2} \right].
\end{array}
\end{equation}
Note that $S_{21}(q) = \overline{S_{12}}(q)$ is a general property of the
Fermion propagator $S$ and can be proved for an arbitrary number of
colors and flavors using only the definition of the inverse Fermion
propagator, \eqref{Sinv}, and properties of the Dirac
gamma matrices.  Whereas only $K$, $L$ and $M$ were used in the derivation
of the gap equation, all these functions are required in evaluating
the Meissner effect.

\begin{figure}[t] \centering
\epsfig{file=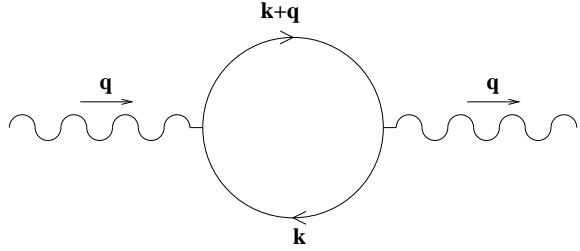,width=3in,bbllx=0,bblly=0,bburx=400,bbury=171}
\caption{One-loop contribution to the Meissner effect.} \label{Meissnerfig} 
\end{figure}
The Meissner effect is the change in the screening of the gluon
propagator induced by the presence of a gap. 
To one loop order, we need to evaluate the gluon propagator
of Fig. \ref{Meissnerfig} using the full fermion propagator
including the gap.  
The result can still be written in the form (\ref{gluonprop})
but now 
\begin{equation}
F(q)=F_0(q)+\delta F(q)\ \ {\rm and} \ \ G(q)=G_0(q)+\delta G(q)
\end{equation}
where $F_0$ and $G_0$ are the $\Delta=0$ functions written
as $F$ and $G$ in (\ref{GF}).  Recall that $G_0$, which
describes Landau damping, vanishes for $q_0\rightarrow 0$.
Because $\delta G$ is nonzero
in the $q_0\rightarrow 0$ limit, the Meissner effect can be
described as giving a mass to the gluons.  
Previous analyses of the Meissner effect have either been
done for two-flavor QCD \cite{Rischke1,CarterDiakonov2} or have 
used simplified estimates \cite{HMSW,ShovkovyWije,EHHS}. 
Our goal is to formulate the correct calculation of 
$\delta F(q)$ and
$\delta G(q)$ in the CFL phase.  Recent work along the same
lines can be found in Ref. \cite{Rischke2}. 

{}From the diagram of Fig. \ref{Meissnerfig}, we obtain the 
gluon polarization
\begin{equation} \label{Pimunu}
\begin{array}{lll}
\Pi^{\mu\nu}_{ab} & = -i g^2 \int {d^4k\over(2\pi)^4} {\rm Tr} & \left[
\Gamma^\mu_a S(k+q) \Gamma^\nu_b S(k) \right] \\
 & = -i g^2 \int {d^4k\over(2\pi)^4} {\rm Tr} & \left[ \gamma^\mu
{\lambda_a\over 2} S_{11}(k+q) \gamma^\nu {\lambda_b\over 2} S_{11}(k) +
\left(\gamma^\mu {\lambda_a\over 2}\right)^T S_{22}(k+q)
\left(\gamma^\nu {\lambda_b\over 2}\right)^T S_{22}(k) \right. \\
 & & \left. - \gamma^\mu {\lambda_a\over 2} S_{12}(k+q) \left(\gamma^\nu
{\lambda_b\over 2}\right)^T S_{21}(k) - \left(\gamma^\mu {\lambda_a\over
2}\right)^T S_{21}(k+q) \gamma^\nu {\lambda_b\over 2} S_{12}(k) \right],
\end{array}
\end{equation}
where the trace is taken over color, flavor, and Dirac indices and all
four elements of the fermion propagator, $S(q)$, have been 
defined previously in \eqsref{S11}{AK}. This polarization amplitude
contains all the one loop contributions to the gluon propagator
including the gap independent contributions, $F_0(q)$ and $G_0(q)$. 
$\Pi^{\mu\nu}_{ab}$ can be written in terms of $F$ and 
$G$ in a simple fashion:
\begin{equation}
\Pi^{\mu\nu}_{ab} = \delta_{ab} \left[ \left( G_0(q) + \delta G(q) \right)
P^{\mu\nu T} + \left( F_0(q) + \delta F(q) \right) P^{\mu\nu L} \right].
\end{equation}
Hence, we only need to compute two components of $\Pi^{\mu\nu}_{ab}$ 
in order to
obtain the functions $\delta F(q)$ and $\delta G(q)$, for example,
$\Pi_{33}^{00}$ and $\Pi_{33}^{11}$.
Because we already know $F_0(q)$ and $G_0(q)$, our goal is to extract
$\delta F(q)$ and
$\delta G(q)$. We are therefore only interested in the
difference $\Pi^{\mu\nu}_{ab}(\Delta \ne 0) - \Pi^{\mu\nu}_{ab}(\Delta =
0)$. Finally, because $\delta F(q)$ and $\delta G(q)$ depend only on
$q_0$ and $|\vec{q}|$, we can choose $\vec{q}$ to lie along the $z$-axis
for simplicity. Keeping all this in mind, we find that (in Euclidean
space)
\begin{eqnarray} \label{dFdG}
\delta F(q) = {q_0^2 + |\vec{q}|^2 \over |\vec{q}|^2} \left(
\Pi^{00}_{33}(\Delta \ne 0) - \Pi^{00}_{33}(\Delta = 0) \right)
\nonumber \\
\delta G(q) = \Pi^{11}_{33}(\Delta \ne 0) - \Pi^{11}_{33}(\Delta = 0).
\end{eqnarray}
Note that (unlike the integrals which arise on the 
right hand side of the gap equation) the integrals which
must be done in evaluating $\Pi(q)$ are ultraviolet divergent,
and therefore sensitive to how they are cutoff at large 
$k_0$ and $k$.  This ultraviolet divergence has nothing 
to do with $\Delta$, and is canceled in our calculation
of $\delta F$ and $\delta G$ by subtracting
the $\Delta=0$ result for $\Pi(q)$.
We have checked that our results for $\delta F$ and $\delta G$ 
are insensitive
to the ultraviolet cutoffs in the integrals.

Looking back at the definition of $\Pi^{\mu\nu}_{ab}$, we can see that
it depends on $\Delta_{1}^{A,S}(k_0)$ and
$\Delta_{2}^{A,S}(k_0)$. We make the same assumptions here
as in our solution of the gap equation, namely that the antiparticle
and sextet contributions can be neglected if $\Delta\ll \mu$
and if one is interested
in physics dominated by particles and holes near the Fermi 
surface.
Before we proceed, let us define the
following notation for the functions $A(q)$ through $M(q)$ defined in
\eqref{AK}: identify the scalar functions multiplying the $P_\pm$
projectors with the appropriate $\pm$ signs, e.g. $A_+(q)$. With this
notation, the dominant contributions to the two polarization amplitudes
we are interested in are:
\begin{equation} \label{Pi0011}
\begin{array}{ll}
\Pi^{00}_{33} = -{i\over 2} g^2 \!\int \!{d^4k\over(2\pi)^4} & \!\left(
1 + \widehat{(k+q)} \cdot \hat{k} \right) \left[ A_-(k+q) A_-(k) -
B_-(k+q) B_-(k) \right. \\  & + 2 C_-(k+q) C_-(k) + E_+(k+q) E_+(k) -
H_+(k+q) H_+(k) \\  & + 2 D_+(k+q) D_+(k) - 2 K_-(k+q) K_-(k) \\
& \left. + 2 L_-(k+q) L_-(k) + 2 M_-(k+q) M_-(k) \right] \\ 
\Pi^{11}_{33} = -{i\over 2} g^2 \!\int \!{d^4k\over(2\pi)^4} & \!\left(
1 + 2 \widehat{(k+q)}^1 \hat{k}^1 - \widehat{(k+q)} \cdot \hat{k}
\right) \left[ A_-(k+q) A_-(k) \right. \\ & - B_-(k+q) B_-(k) + 2
C_-(k+q) C_-(k) + E_+(k+q) E_+(k) \\ & - H_+(k+q) H_+(k) + 2 D_+(k+q)
D_+(k) + 2 K_-(k+q) K_-(k) \\ & \left. - 2 L_-(k+q) L_-(k) - 2 M_-(k+q)
M_-(k) \right].
\end{array}
\end{equation}
In any one gauge, i.e. for a particular choice of $\xi$, our task
is now clear. We first calculate $\Delta(k_0)$ with 
$\delta F(q)=\delta G(q)=0$,
as described in the body of the paper.  We 
must then use (\ref{Pi0011})
to evaluate $\delta F(q)$ and $\delta G(q)$ given by \eqref{dFdG}. 
As in the calculation of $\Delta$, 
we can do all angular integrals analytically and evaluate
the double integral over $k_0$ and $|{\vec k}|$ 
numerically.
We must then re-evaluate $\Delta(k_0)$ 
with the new gluon propagator, modified by the addition
of $\delta F(q)$ and $\delta G(q)$.  We must then iterate 
this procedure, calculating $\delta F(q)$ and $\delta G(q)$ and
then recalculating $\Delta(k_0)$ repeatedly,
until all results
have converged.
We have not carried this program to completion. However, preliminary
numerical investigation suggests that, in agreement with
arguments and estimates made by 
others \cite{Son,SW,HMSW,HsuSchwetz,ShovkovyWije,EHHS}, the change in $\Delta$
arising from the inclusion of $\delta F$ and $\delta G$ is small.
In particular, it
appears to be much smaller than the change in $\Delta$ which arises
if one changes gauge from $\xi=-1$ to $\xi=0$ to $\xi=1$.  Perhaps at
some extremely small $g$, the influence of the Meissner
effect on the gap could be larger than the influence of
the neglected physics whose absence we diagnose via
the gauge dependence of our results.  At any $g$ at which
we have been able to obtain numerical  results, however, the Meissner
effect is insignificant relative to that which is missing.

\end{document}